\documentclass[12pt]{article}
\RequirePackage{amsbsy}
\RequirePackage{amssymb}
\RequirePackage{mathrsfs}
\RequirePackage{epsfig}

\def\nn{\nonumber}
\def\a'{a^{\prime}}
\def\f'{f^{\prime}}
\def\h'{h^{\prime}}

\begin{document}
\begin{titlepage}

\title{Gravitationally coupled magnetic monopole and conformal symmetry breaking}

\author{$^1$Ariel Edery\thanks{Email: aedery@ubishops.ca}$\quad,\quad^2$Luca Fabbri\,\thanks{Email:
lucafabbri@bo.infn.it} $\quad,\quad^3$M.B. Paranjape\thanks{Email:
paranj@lps.umontreal.ca}\\\\
{\small\it $^1$Physics Department, Bishop's University}\\
{\small\it 2600 College Street, Sherbrooke, Qu\'{e}bec, Canada
J1M~0C8} \\{\small\it $^2$Theory Group, INFN Section of Bologna,
Department of physics }\\{\small\it- University of Bologna, Via
Irnerio 46, C.A.P. 40126, Bologna, ITALY }\\{\small\it $^3$ Groupe
de physique des particules, Universit\'e de Montr\'eal,}\\{\small\it
C.P. 6128, succ. centre-ville, Montr\'eal, Qu\'ebec, CANADA  H3C
3J7} }
\date{}
\maketitle
\begin{abstract}
We consider a Georgi-Glashow model conformally coupled to gravity.
The conformally invariant action includes a triplet of scalar fields
and $SO(3)$ non-Abelian gauge fields. However, the usual mass term
$\mu^2\,\phi^2$ is forbidden by the symmetry and this role is now
played by the conformal coupling of the Ricci scalar to the scalar
fields. Spontaneous symmetry breaking occurs via gravitation. The
spherically symmetric solutions correspond to localized solitons
(magnetic monopoles) in asymptotically anti-de-Sitter (AdS)
space-time and the metric outside the core of the monopole is found
to be Schwarzschild-AdS. Though conformal symmetry excludes the
Einstein-Hilbert term in the original action, it emerges in the
effective action after spontaneous symmetry breaking and dominates
the low-energy/long-distance regime outside the core of the
monopole.
\end{abstract}

\parskip=.5cm
\end{titlepage}
\newpage
\section{Introduction}
\setlength{\parindent}{0em} \setlength{\parskip}{0.6em} This paper
is based on an invited talk given at Theory Canada IV in June, 2008
at the Centre de Recherche Math\'{e}matique, Universit\'{e} de
Montr\'{e}al. The talk discussed a previously published paper
\cite{manu}. In these proceedings we summarize the paper and also
highlight various points in the conclusion section that were not
discussed in much detail in \cite{manu}.

A conformal transformation is a rescaling of the metric at every
coordinate $x$. Such a transformation alters the curvature of
spacetime but in such a way that the light cones, and hence the
causal structure of spacetime, is preserved. This renders conformal
symmetry of potential interest. However, before one can incorporate
such a symmetry there are two things to consider. The introduction
of a scale, such as a mass term in the action, breaks the symmetry.
Nature is not conformally invariant and one therefore needs a
mechanism that breaks the original symmetry. Secondly, though some
well-known massless field equations are conformally invariant, such
as Maxwell's equations in 3+1 dimensions, others such as the
Klein-Gordon equation for a massless scalar field are not. One must
add an extra term to the usual scalar field kinetic term to obtain
an invariant equation. This extra term is the conformal coupling of
the Ricci scalar to the scalar field. Interestingly, it plays a dual
role. Besides rendering the action for the scalar field conformally
invariant it also acts as part of a symmetry breaking potential.

Though conformal symmetry forbids the inclusion of the
Einstein-Hilbert term in the original action, we find that the
effective action after symmetry breaking is dominated by Einstein
gravity (General Relativity) in the long distance or low energy
regime. The particular model we consider is a Georgi-Glashow model
\cite{gg} conformally coupled to gravity. As in the usual
Georgi-Glashow model we include a triplet of scalar fields and SO(3)
non-abelian gauge fields. However, the usual mass term
$\mu^2\,|\phi| ^2$ is forbidden by the conformal symmetry and is
replaced by $R\,|\phi|^2/6$ i.e. the Ricci scalar conformally
coupled to the scalar fields. Our solutions have a non-singular core
representing localized soliton solutions of the magnetic monopole
type in an asymptotically anti-de-Sitter (AdS) space-time. The
metric for the spherically symmetric case is Schwarzschild-AdS
outside the core of the monopole and we obtain numerical solutions
for the scalar fields, the gauge fields and the metric both inside
and outside the core of the monopole. Conformally coupled
matter/gravity and also spontaneous breaking of conformal invariance
has been considered by Demir, by Odintsov and by Mannheim and
Kazanas \cite{domk}. The latter work is the most relevant to ours.
They consider the conformally coupled scalar/gravitational field
equations and find for a negative scalar field self-coupling that a
mass scale is generated in the theory and that the metric
corresponds to de-Sitter space-time. Other work of interest for the
case of Einstein gravity or string inspired gravity with Yang-Mills
and Higgs fields can be found in \cite{hbfrghs}.
\section{Metric, scalar and gauge fields in a conformally invariant action}
We construct an action which is both conformally invariant and
generally covariant (our metric signature is $(+,-,-,-)$). A
conformal transformation is a scaling of the metric at every
spacetime point:
\begin{eqnarray}
\nn
g'_{\mu \nu}=\Omega^{2}g_{\mu \nu}
\end{eqnarray}
where $\Omega=\Omega(x)$ is a smooth positive function. The
gravitational part of the action is given by the Weyl tensor squared
instead of the Einstein-Hilbert term:
\begin{eqnarray}
\nn S=\int\sqrt{|g|}C_{\beta \mu \nu \rho}C^{\beta \mu \nu
\rho}d^{4}x \equiv \int\sqrt{|g|}C^{2}d^{4}x\,.
\end{eqnarray}
This is the unique gravitational action constructed using only the
metric, which is invariant under (local) conformal and coordinate
transformations. We consider a gauge potential as $n$ $4$-vectors
$A_{\mu}^{a}$. The field strength is then given by
\begin{equation}
F_{\mu \nu}^{a}=\nabla_{\mu}A_{\nu}^{a}-\nabla_{\nu}A_{\mu}^{a}
+C^{a}_{\phantom{a}bc}A_{\mu}^{b}A_{\nu}^{c}
\end{equation}
where $C^{a}_{\phantom{a}bc}$ are the structure constants of the Lie
algebra of the gauge group. The conformal transformation for the
gauge field is defined as $A_{\mu}^{'a}=A_{\mu}^{a}$ and this yields
$F_{\mu \nu}^{'a}=F_{\mu \nu}^{a}$. The Higgs boson is defined as
$n$ scalars $\phi^{a}$. The gauge covariant derivative of Higgs
bosons is given by
\begin{equation}
D_{\mu}\phi^{a}=\nabla_{\mu}\phi^{a}+C^{a}_{\phantom{a}bc}A_{\mu}^{b}\phi^{c}.
\end{equation}
The action for the Higgs bosons in interaction with the gauge field
is given by the standard form (with implicit summation over $k$
assumed)
\begin{equation}
S=\int\sqrt{|g|}\left(
D_{\mu}\phi^{k}D^{\mu}\phi_{k}-\frac{1}{4e^{2}}F_{\mu \nu}^{k}F^{\mu
\nu}_k\right)d^{4}x. \label{act}
\end{equation}
A conformal transformation for the Higgs is defined as
$\phi'^{a}=\Omega^{-1}\phi^{a}\label{h}$ and the term
$D_{\mu}\phi^{k}\,D^{\mu}\phi_{k}$ in \ref{act} is not conformally
invariant. However, by adding the extra term
$\frac{1}{6}R\phi^{k}\phi_{k}$ i.e. the conformal coupling of the
Ricci scalar to the scalar fields, the action is made conformally
invariant. We are also free to add a $\phi^4$ self-coupling term
$\lambda^{2}(\phi^{k}\phi_{k})^{2}$, which is conformally invariant
by itself. This will be part of a symmetry breaking potential. The
complete conformally invariant action is then given by \cite{manu}
\begin{equation}
S=\int\sqrt{|g|}\,\,\Big(\,C^{2}-\frac{1}{4e^{2}}F^{2}+(D\phi)^{2}+\frac{1}{6}R\phi^{2}-\lambda^{2}\phi^{4}\,\Big)\,\,d^{4}x
\label{eqy}\end{equation} where $C^2= C_{\beta \mu \nu \rho}C^{\beta
\mu \nu \rho}$, $F^2= F_{\mu \nu}^{k}F^{\mu \nu}_k$ and
$\phi^2=\phi^{k}\phi_{k}$ with implicit summation over $k$. The
above is a conformally invariant action that includes the
interaction between the metric, the gauge fields and the Higgs
bosons. The quantity $e$ is the gauge coupling constant.
\subsection{Symmetry breaking via gravitation}
In (\ref{eqy}) the last two terms can be seen as a non-minimal
potential for the scalar field
\begin{equation}
V(\phi)=-\frac{1}{6}R\phi^{2}+\lambda^{2}\phi^{4}.
\end{equation}
Note that the usual mass term $\mu^2\,\phi^2$ is absent from the
potential. The term with $\phi^{2}$ is proportional to the Ricci
scalar $R$, which is not the mass of Higgs field, nor a constant.
The above potential can provide for spontaneous breaking of the
conformal and gauge symmetry. We consider the case of positive
quartic scalar self-coupling, $\lambda^2>0$ which is required for
stability in flat backgrounds. Spontaneous symmetry breaking can
occur, for example, when the scalar curvature is a constant positive
value, namely, in AdS space-times. It is too great of a restriction
to impose that $R$ is a positive constant everywhere and we make the
\emph{Ansatz} that the AdS geometry is only asymptotic, that is the
Ricci scalar goes to a positive constant only in the limit for which
the independent variables are at the boundary of the particular
system of coordinates we will consider. In asymptotically AdS
space-times, then, we will set the numerical value of the Ricci
scalar at the boundary of that specific system of coordinates to be
equal to $12\lambda^{2}\,v^{2}$, with $v$ an arbitrary constant.  In
this case, $\phi^{2}=v^{2}$ is the solution which asymptotically
gives the spontaneous breakdown of the gauge symmetry, giving mass
to some of the gauge fields via the Higgs mechanism. The conformal
symmetry is therefore broken in the process.

\section{Spherically symmetric construction: metric, gauge and
scalar fields} In the Georgi-Glashow model the basic assumption is
that we have a triplet of Higgs bosons and a triplet of gauge fields
for which the symmetry group is $SO(3)$. The geometrical
configuration of the spacetime we want to study is chosen to be
stationary and spherically symmetric, i.e. the distribution of
energy does not depend on time and it is isotropically distributed
around the origin, hence, the symmetry group of the spatial
isometries is also $SO(3)$.

This structure for spatial isometries allows us to consider $3$
linearly independent $4$-dimensional Killing vectors in spherical
coordinates $(t,r,\theta,\varphi)$, given as
follows:$\xi_{(1)}\!=\!(0,0,\cos{\varphi},-\sin{\varphi}\cot{\theta}),
\xi_{(2)}\!=\!(0,0,-\sin{\varphi},-\cos{\varphi}\cot{\theta}),
\\\xi_{(3)}=(0,0,0,1)$, for which
$[\xi_{(i)},\xi_{(j)}]=\varepsilon_{ijk}\xi_{(k)}$, the correct
representation of the symmetry group for $SO(3)$.

The line element in spherical coordinates has to be isotropic, that is there exists a system of coordinates in which it has the form
\begin{equation}
\nn
ds^{2}=A(r)dt^{2}-B(r)dr^{2}-r^{2}(d\theta^{2}+\sin^{2}{\theta}d\varphi^{2})
\end{equation}
It has been noted by Mannheim and Kazanas (\cite{km}) that, via a sequence of coordinate and conformal transformations,
one can always bring the general static, spherically symmetric metric to a form in which $A=1/B$.
Thus it is justified to set $A=1/B=1+h(r)$ for a smooth function $h(r)$, so that
\begin{equation}
ds^{2}=(1+h(r))dt^{2}-\left(\frac{1}{1+h(r)}\right)dr^{2}-r^{2}(d\theta^{2}+\sin^{2}{\theta}d\varphi^{2}).
\label{metric}
\end{equation}
Given this line element, we can compute the curvature tensors (see
\cite{manu}). Asymptotically the spacetime is AdS where the Ricci
scalar is given by
\begin{equation}
\lim_{r \to \infty}R(r)=12\lambda^{2}\,v^{2} \label{bound}
\end{equation}

We note that $SO(3)$ is at the same time the spatial and internal
(gauge) group of symmetries. The three gauge fields can then be
expressed in terms of the three above Killing vectors (using
spherical coordinates): $A_{i}^{\mu}= q(r^2) \,\xi_{(i)}^{\mu}$
where $q(r^2)$ is a function of $r^2$. For the Higgs field we obtain
$\phi^{a}=f(r^{2})\frac{r^{a}}{r}$ where $f(r^2)$ in another
function of $r^2$. Instead of working with $q(r^2)$ it is more
convenient to work with the function $a(r^2)$ defined by
\begin{equation}
1+r^{2}q(r^{2})=a(r^{2})\,. \label{supplem}
\end{equation}
The field strength and the gauge covariant derivatives can now be
written down explicitly for this theory. The equations of motion for
each field are obtained by varying the effective action with respect
to the radial functions $h(r),f(r),a(r)$. This yields the equations
\cite{manu}
\begin{equation}
(rh)''''=\frac{r}{2}(2f'^{2}-ff'')+\frac{3a'^{2}}{2re^{2}}\label{e1}
\label{f1}\end{equation}
\begin{equation}
((1+h)r^{2}f')'=f(2a^{2}+r^{2}(2\lambda^{2}f^{2}-\frac{(r^{2}h)''}{6r^{2}}))\label{e2}
\label{f2}\end{equation}
\begin{equation}
((1+h)a')'=a(2f^{2}e^{2}+\frac{a^{2}-1}{r^{2}})\label{e3} \label{f3}
\label{f3}\end{equation} where the prime is the derivative with
respect to $r$.
\begin{figure}
\begin{center}
\includegraphics[width=12cm]{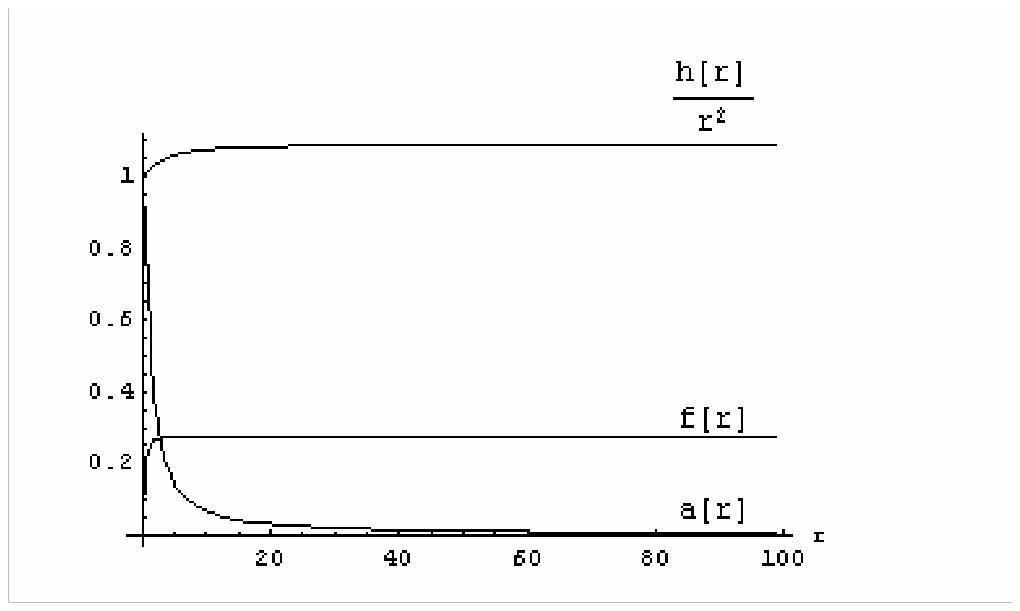}
\label{fig: Plot of the Solutions} \caption{Plot of $a(r)$, $f(r)$
and $h(r)/r^2$ for $e=1$.}
\end{center}
\end{figure}
\subsection{Results: analytical and numerical solutions}
Outside a core region the vanishing N\"other's density allows a
non-trivial solution $a=0$. This is non-trivial because there is a
long range monopole like magnetic field. It is possible to solve the
field equations exactly for this external solution. A solution with
spontaneous symmetry breaking of the conformal symmetry requires
that $f$ reaches a nonzero arbitrary constant $v$ asymptotically. We
consider the analytical solution with $f=v$ identically. The only
possible solution for the metric is given as
$h=\frac{k}{r}+\lambda^{2}v^{2}r^{2}$. One solution for the exterior
region is then $a(r)=0$, $f(r)=v$ and $
h(r)=\frac{-\beta}{r}+\lambda^{2}v^{2}r^{2}$ for any value of the
free parameter $\beta$ and for any positive value of the parameter
$v$. The quadratic term, which is dominant at large distances, gives
rise to a constant, positive Ricci scalar for the positive
$\lambda^{2}v^{2}$. Hence, the space-time is asymptotically
anti-de-Sitter. Other analytical solutions are discussed in
\cite{manu}. Outside the core region we therefore obtain
Schwarzschild-AdS spacetime. The behaviour at shorter distances
inside the core requires numerical solutions (see Figure 1). In the
core it is clearly seen that Einstein gravity no longer dominates.

The solution $a=0$ means, from (\ref{supplem}), that $q(r^{2})$
behaves as $-1/r^{2}$. This gives rise to a non-zero purely radial
magnetic field strength $
B^{(k)r}=\frac{\phi^{(k)}}{|\phi|}\frac{1}{r^{2}}$ that is
proportional to the Higgs field, in the internal space \cite{manu}.
Locally we can choose (apart from the origin) $\vec{\phi}= (0,0, v)$
and then the third component of the gauge field is
$B^{(3)r}=\frac{1}{r^{2}}$.  This third component corresponds to the
$SO(2)$ unbroken symmetry; hence, it can be identified with the
Abelian magnetic field.

We can look for numerical solutions to the field equations \ref{f1},
\ref{f2} and \ref{f3}. Such a numerical solution includes the
behaviour of the core region (i.e. the magnetic monopole) as well as
the external region. The field equations need eight initial
conditions to be given at the origin which specify a unique
solution. A simple program finds the solutions of the type that we
are looking for i.e. where asymptotically $f=v$ and $h= \lambda^2
\,v^2$. A set of initial conditions can be chosen to give the
solution presented in Figure 1. The most important point about the
numerical solution is that the core is non-singular. This is where
the magnetic monopole resides (the region in the neighborhood of
$r=0$ where $a(r)$ is significant). The spacetime is AdS
asymptotically with Schwarzschild in the intermediate region.
However, it is easy to see that the Schwarzschild solution does not
extend to $r=0$ in the core region as there is no Schwarzschild
singularity there.

\section{Conclusion: discussion and future work}

We have focused thus far on the broken sector where the exterior
solution is Schwarzschild AdS. There is also a non-broken sector
where $|\phi|=0$ asymptotically. This yields the more general vacuum
solution originally found by Mannheim and Kazanas \cite{km}. The
hallmark of that solution is the presence of a linear potential
$\gamma\,r$ ($\gamma$ is an integration constant) besides the usual
$1/r$ Newtonian potential. This arises because in the non-broken
sector one actually has a fourth order gravity theory due to the
presence of the conformally invariant $C^2$ term, the Weyl tensor
squared. Unlike General Relativity, this yields a fourth order
poisson equation which has a linear potential as part of its
solution. What is interesting is that spontaneous symmetry breaking
eliminates altogether this long range linear potential even though
the $C^2$ term is still present in the Lagrangian. The spontaneous
symmetry breaking acts as a ``switch" from fourth order gravity to
second order gravity at large distances. Outside the core, one finds
the Schwarzschild metric only (in an AdS background). Let us
investigate this further.

Einstein gravity (General Relativity) is not present in the original
action. In fact, the original Lagrangian looks like fourth order
Weyl gravity with matter coupled to the Ricci scalar. However
Einstein gravity is {\it induced} via spontaneous symmetry breaking.
This can be seen by expanding the action \ref{eqy} about $|\phi|=v$.
One of the terms in the expansion is $(1/6)\,v^2 R$. This is nothing
other than the Einstein-Hilbert term with $v^2/6$ identified with
$1/(16\,\pi\,G)$ where $G$ is Newton's gravitational constant.
Another term in the expansion is $\lambda^2 \,v^4$ arising from
$(1/6)R\phi^2 -\lambda^2\phi^4$ where $R=12\,\lambda^2\,v^2$ in AdS
spacetime. This acts as a cosmological constant. In the broken
sector, where $|\phi|=v$ asymptotically, Einstein gravity (General
Relativity) with a cosmological constant dominates over the
higher-derivative $C^2$ term in the low energy or long distance
regime. This is why there is no linear term in the potential.
Besides inducing Einstein gravity, an important and interesting
feature of the solution is that there is no singularity at $r=0$.
The numerical results show a regular magnetic monopole solution in
the interior. The core region is clearly not dictated by the
Schwarzschild metric. This makes sense because in the core region it
is fourth order gravity that now dominates over the Einstein-Hilbert
term. The roles switch in the interior.

Inducing Einstein gravity starting with a different theory may be
important in general. At present, theoretical physics is faced with
the cosmological constant problem. Since gravity couples to energy
it should couple to vacuum energy. This should lead to a very large
cosmological constant. However, this is not observed. This is not a
quantum gravity problem though the two problems may be ultimately
related. There is no reason to doubt the existence of quantum vacuum
fluctuations since they are fundamental, stemming from the
uncertainty principle. Yet, when we include matter vacuum
fluctuations in the calculation of the expectation value of the
energy-momentum tensor we obtain an absurdly large cosmological
constant. GR is a very successful theory and has been tested
experimentally in both the weak and strong field regimes. One
possibility then is that GR is an induced theory and emerges after
matter vacuum fluctuations are included in a fundamentally different
theory. What we have shown in this paper is that the idea of induced
gravity, at least at the classical level, is a realizable project.
Moreover, it is not realized in an arbitrary fashion. The original
theory is motivated by symmetry, in our case conformal symmetry.
Contact with nature is then achieved via spontaneous symmetry
breaking. Again, we see the powerful and fruitful role that both
symmetries and symmetry-breaking play.

For future work, it would be interesting and important to add
quantum corrections to the action. These contribute
higher-derivative geometrical terms such as the Ricci tensor squared
$R_{\alpha\beta}R^{\alpha\beta}$ and Ricci scalar squared $R^2$ to
the gravitational part of the action \cite{Mukhanov}. The action is
then no longer conformally invariant and introduces a fundamental
length scale into the theory. This would ultimately fix the value
$v$ for the expectation value of the scalar field. As already
discussed, the value of $v$ is related to Newton's constant G, a
cosmological constant and massive scalars. These will then have a
non-trivial relation to each other that includes Planck's constant.
This should be interesting.

How will adding quantum corrections affect the equations of motion?
The corrections are higher-derivative terms and should not affect
the long distance regime in the broken sector (after spontaneous
symmetry breaking). Einstein gravity with cosmological constant
should still dominate in this regime but now with parameters fixed
by the introduction of a length scale. What will definitely change
is the numerical solution in the interior, the core region, since
the new higher derivative terms affect this high-energy regime. A
question of interest is whether the magnetic monopole solution will
remain regular at $r=0$.

\section*{Acknowledgments}
 MB and AE acknowledge support from the Natural Sciences and Research Council of Canada (NSERC) and
 LF acknowledges financial support from INFN-Bologna.


\end{document}